# Molten flux growth of single crystals of quasi-1D hexagonal chalcogenide BaTiS$_3$

Huandong Chen[1], Shantanu Singh[1], Hongyan Mei[2], Guodong Ren[3], Boyang Zhao[1], Mythilli Surendran[1,4], Yan-Ting Wang[1], Rohan Mishra[5,3], Mikhail A. Kats[2,6], Jayakanth Ravichandran[1,4,7*]

[1]Mork Family Department of Chemical Engineering and Materials Science, University of Southern California, Los Angeles CA, USA
[2]Department of Electrical and Computer Engineering, University of Wisconsin-Madison, Madison WI, USA
[3]Institute of Materials Science & Engineering, Washington University in St. Louis, St. Louis MO, USA
[4]Core Center for Excellence in Nano Imaging, University of Southern California, Los Angeles CA, USA
[5]Department of Mechanical Engineering and Materials Science, Washington University in St. Louis, St. Louis MO, USA
[6]Department of Materials Science and Engineering, University of Wisconsin-Madison, Madison WI, USA
[7]Ming Hsieh Department of Electrical and Computer Engineering, University of Southern California, Los Angeles CA, USA

*Email: j.ravichandran@usc.edu

**Abstract**

BaTiS$_3$, a quasi-1D complex chalcogenide, has gathered considerable scientific and technological interest due to its giant optical anisotropy and electronic phase transitions. However, the synthesis of high-quality BaTiS$_3$ crystals, particularly those featuring crystal sizes of millimeters or larger, remains a challenge. Here, we investigate the growth of BaTiS$_3$ crystals utilizing a molten salt flux of either potassium iodide, or a mixture of barium chloride and barium iodide. The crystals obtained through this method exhibit a substantial increase in volume compared to those synthesized *via* the chemical vapor transport method, while preserving their intrinsic optical and electronic properties. Our flux growth method provides a promising route towards the production of high-quality, large-scale single crystals of BaTiS$_3$, which will greatly facilitate advanced characterizations of BaTiS$_3$ and its practical applications that require large crystal dimensions. Additionally, our approach offers an alternative synthetic route for other emerging complex chalcogenides.

## Introduction

Complex chalcogenides, highlighted by chalcogenide perovskites, have gathered considerable research interest recently due to their exciting electronic and optical properties[1-3]. Among them, the quasi-1D hexagonal chalcogenide, BaTiS$_3$, is a small bandgap semiconductor[2] that has shown pronounced optical anisotropy at mid-wave and long-wave infrared energies[2,4], making it particularly attractive for polarization-selective infrared optics and optoelectronic devices[5-7]. Furthermore, the recent discovery of charge density wave (CDW) electronic phase transitions[8] in single crystalline BaTiS$_3$, along with the successful demonstration of neuronal device functionalities using this material[9], opens up compelling pathway for the development of electronic devices based on this novel phase-change material. However, despite the exciting physical phenomena and the promising multifunctionality of BaTiS$_3$, its practical applications, especially for infrared optics, are largely limited by the attainable crystal sizes. For instance, the fabrication of BaTiS$_3$-based optical components, such as waveplates, necessitates crystals with well-defined optical axes and lateral dimensions of several millimeters or larger. Another compelling motivation for obtaining sizable and high-quality single crystals of BaTiS$_3$ arises from the urgent need of advanced characterization techniques, including neutron-based diffraction and scattering, to probe the underlying electronic phase transition mechanism. Many of these techniques require large crystal dimensions or volume[10,11] for meaningful analysis.

To date, the synthesis of BaTiS$_3$ single crystals has exclusively relied on the iodine-assisted chemical vapor transport (CVT) method[2,12], where the size and morphology of crystals have certain limitations. For example, the quasi-1D BaTiS$_3$ crystals produced *via* this method tend to form clusters of "thin needles" out of the powder due to the excessive number of nucleation sites (powder) during the growth process. Consequently, their dimensions perpendicular to the chain axis (both width and thickness) typically remain below 50 µm[12]. This is in stark contrast to the successful applications of CVT in producing various large-sized 2D transition metal dichalcogenide single crystals[13,14]. An alternative, and widely used synthesis technique is the high temperature solution or molten flux growth method. This approach has been employed in obtaining various high-quality crystals with lateral dimensions of several millimeters or larger, benefiting from a well-controlled small number of nucleation sites and a slow cooling rate[15,16]. As an example, millimeter-sized single crystals of the isostructural hexagonal chalcogenide $BaVS_3$ have already been obtained using either tellurium flux[17] or barium chloride flux[18]. Moreover, the development of flux growth can further advance other solution-based synthesis techniques such as solution Bridgeman, towards the realization of wafer-scale complex chalcogenide single crystals.

In this work, we present the first comprehensive study of single crystalline $BaTiS_3$ growth *via* the flux method, using either potassium iodide (KI) or a mixture of barium chloride ($BaCl_2$) and barium iodide ($BaI_2$) as the salt flux. This KI flux approach yielded crystals of dimensions up to a centimeter in length and 500 µm in both width and thickness, while the $BaCl_2$-$BaI_2$ flux method produced thick (up to 200 µm) and plate-like $BaTiS_3$ crystals with (001)-orientation out-of-plane, both of which are significantly larger in volume compared to those synthesized through the conventional CVT technique[2,19]. The high quality of the flux-grown crystals was validated through chemical and structural characterization conducted at room temperature. Optical anisotropy with a giant birefringence of up to 0.8 in the mid-infrared region and a substantial dichroic window from 1 to 4 µm were revealed in a flux-grown $BaTiS_3$ single crystal, through polarization-dependent infrared spectroscopy and ellipsometry analysis. Moreover, the electronic phase transitions were characterized by our combined temperature-dependent structural and transport measurements, the results of which are consistent with those observed in CVT-grown samples with a subtle shift in transition temperatures. Our study provides a feasible route for synthesizing high-quality, large BaTiS₃ single crystals using the flux growth method, which not only paves the way for practical device applications and advanced characterization of BaTiS₃, but also sheds light on the development of high-temperature solution-based synthesis of other complex chalcogenide materials.

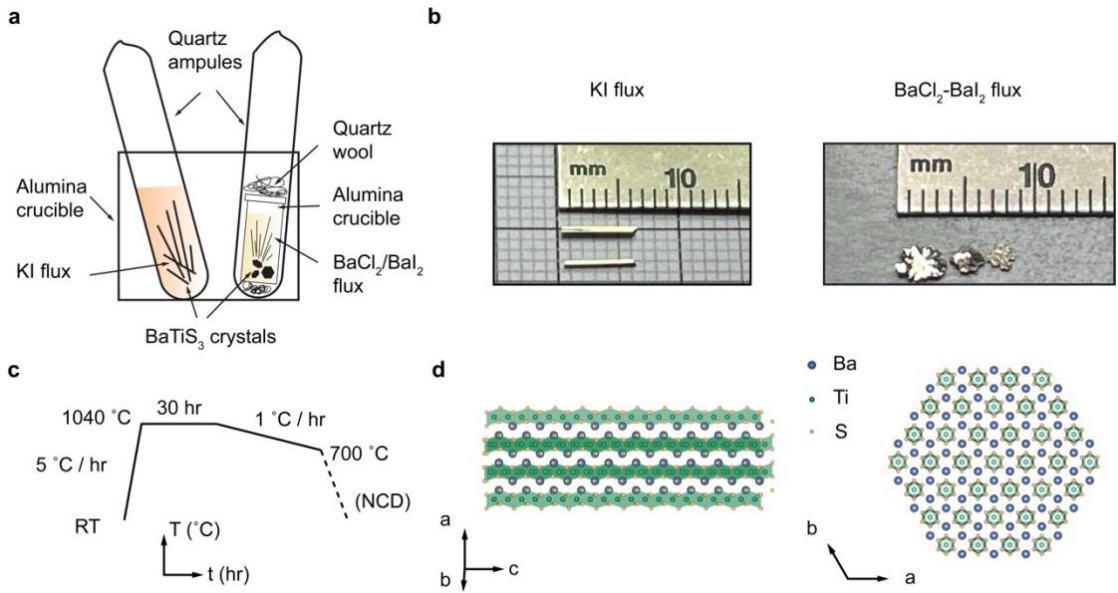

**Figure 1 Molten flux growth of BaTiS₃ single crystals.** (**a**) Schematic illustration of molten flux growth of BaTiS₃ crystals using a vertical geometry. (**b**) Optical image of representative BaTiS₃ crystals grown with KI and BaCl₂-BaI₂ fluxes, respectively. KI flux growth yielded crystals with ~ 6 mm length and half a millimeter in both width and thickness. BaTiS₃ thick-plate-like crystals with 3 mm lateral dimensions and up to 250 µm thickness were obtained from BaCl₂-BaI₂ flux growth. (**c**) Temperature profile for flux growth of BaTiS₃. (**d**) BaTiS₃ crystal structure viewed towards the *a-c* plane and along the *c*-axis.

## Molten flux growth of BaTiS₃ crystals

Polycrystalline BaTiS₃ precursor powders were first prepared by conventional solid-state reaction at 1040°C in a sealed quartz ampoule, and then the as-grown powders were annealed with excess sulfur pieces at 650°C to heal the sulfur vacancies. Similar annealing process has been adopted to guarantee the sulfur stoichiometry when synthesizing BaVS$_3$ crystals[20], whose magnetic properties are sensitive to sulfur vacancies. Single crystals of BaTiS$_3$ were grown using either KI or a BaCl$_2$-BaI$_2$ mixture as the molten flux, as illustrated in Figure 1a. For flux growth using KI, a mixture of the pre-synthesized BaTiS$_3$ powder, pre-dried KI powder (BaTiS$_3$: KI atomic ratio ~ 1:100), and sulfur pieces were placed directly in a sealed quartz ampoule; while for BaCl$_2$-BaI$_2$-based growth, a pre-mixture of BaCl$_2$ and BaI$_2$ was first loaded in a small aluminum crucible with the same BaTiS$_3$ powder (BaTiS$_3$ : BaCl$_2$-BaI$_2$ mixture atomic ratio ~ 1:7) and excess sulfur, and then sealed in a straight quartz ampoule, in order to minimize the corrosion of quartz caused by BaCl$_2$.

The crystal growth procedures remain the same for both fluxes, where the mixture were first heated to 1040°C to fully dissolve all the components and then slowly cooled to 700°C at a rate of 1°C/hr to allow the nucleation and continuous growth of BaTiS$_3$ crystals, after which the furnace was turned off for natural cooling, as illustrated in Figure 1a and 1c. After cooling down to room temperature, as-grown BaTiS$_3$ crystals were extracted by washing off excess salt flux using DI water. BaTiS$_3$ crystals obtained from KI-flux growths typically feature lengths of several millimeters and both widths and thicknesses of up to 500 µm, whose dimensions perpendicular to the chain-axis (along *a*- and *b*- axis) are substantially larger than those of CVT-grown crystals[12], as shown in the optical image (Figure 1b, left). This dominant "thick needle-type" morphology of BaTiS$_3$ with hexagonal cross section is consistent with its quasi-1D crystal structure and a hexagonal symmetry, as illustrated in Figure 1d. Notably, a different "thick plate-like" crystal morphology with *a*- and *b*- axes in-plane (up to several millimeters laterally and 250 µm in thickness) was obtained from growths using BaCl$_2$-BaI$_2$ flux (Figure 1b, right), alongside with regular thin needle-like morphologies. Synthesis of such BaTS$_3$ crystals with *c*-axis out-of-plane has already been reported previously using the conventional CVT method[19]. However, the CVT grown crystals have much smaller sizes, particularly in thickness.

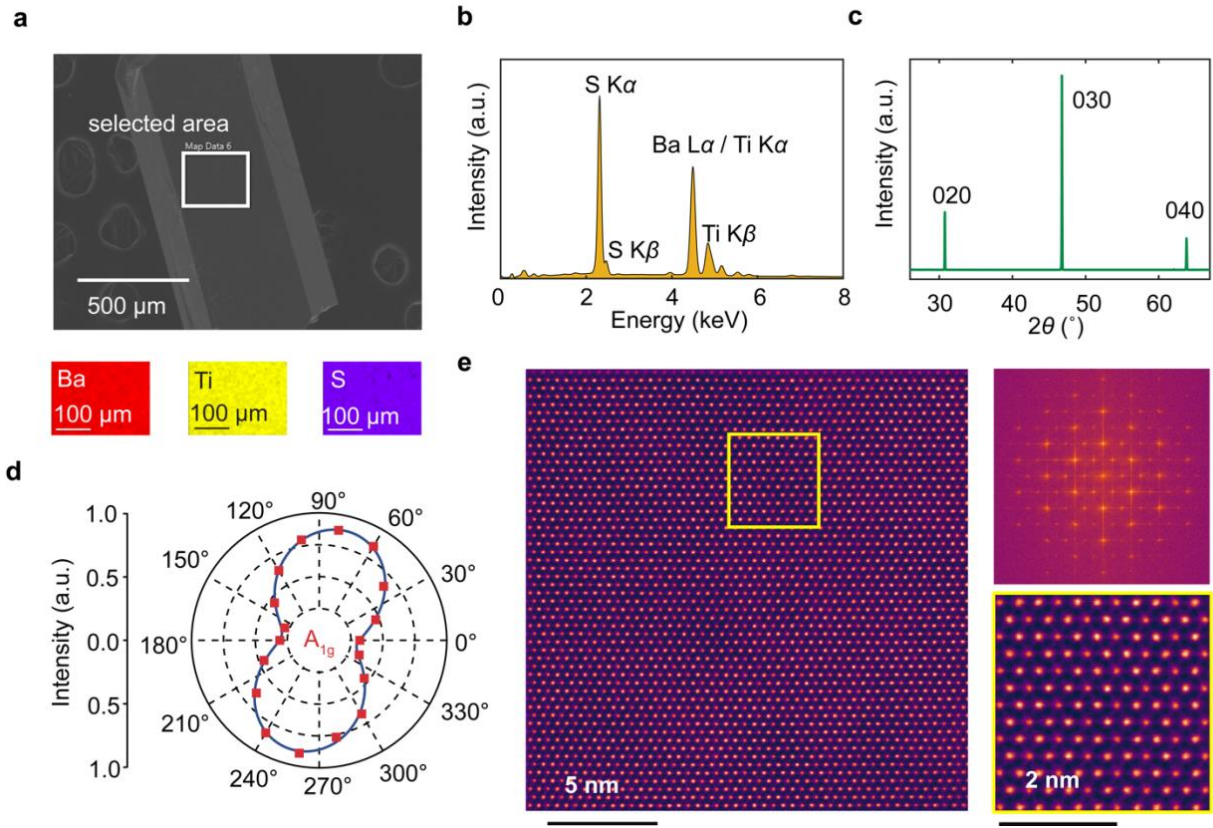

**Figure 2 Room temperature chemical, and structural characterization of flux-grown BaTiS$_3$.** (**a**) SEM image of a representative thick BaTiS$_3$ crystal grown by KI flux. EDS mapping of barium (red), titanium (yellow) and sulfur (purple) elements in the selected area is shown as the inset. (**b**) EDS spectrum of flux-grown BaTiS$_3$ as in Figure 2(**a**), showing Ba : Ti : S ratio as 1 : 1.02 : 2.96. (**c**) Out-of-plane XRD scan of a thick BaTiS$_3$ needle-like crystal grown by KI flux. (**d**) Intensity of A$_{1g}$ Raman mode of BaTiS$_3$ crystal at ~ 380 cm$^{-1}$ with different excitation laser polarizations plotted in polar coordinates. The blue line is the fitted curve. (**e**) Atomic-resolution HAADF-STEM image of a KI flux-grown BaTiS$_3$ crystal viewed along the *c*-axis and the corresponding FFT pattern. Higher magnification HAADF-STEM image of BaTiS$_3$ acquired from the region highlighted with yellow box is illustrated at the bottom right.

The selection of either KI or BaCl$_2$-BaI$_2$ mixture as the flux material for BaTiS$_3$ growth in this work over many other halogen salts and metal fluxes was based on several important considerations as described below, in addition to its capability of dissolving BaTiS$_3$ powder at elevated temperatures. First, KI flux features a relatively low melting point of ~ 681°C, which allows the growth to be performed at low temperatures[21] and enables a large process window for crystal nucleation and continuous growth. Similarly, the melting temperature of the $BaCl_2$-$BaI_2$ mixture is expected to be much reduced compared to pure $BaCl_2$ ($T_m$ ~ 962°C), due to the addition of the low melting temperature component $BaI_2$ ($T_m$ ~ 711°C). Second, compared to other commonly used halogen salts such as $BaCl_2$[18,22], KI is chemically less aggressive and does not require the use of doubly sealed quartz ampoules or special crucible materials such as alumina to protect the furnace and the furnace tubes. Hence, the overall growth process is significantly simplified. As for the $BaCl_2$-$BaI_2$ mixture, we applied an alumina crucible as the inner container and a second outer quartz sealing, in order to minimize the corrosion of quartz ampoule wall, which reduces the chance of leaking or explosion during the growth. Finally, unlike many metal fluxes that require high temperature centrifuging process for crystal separation[15,23], the excess KI or $BaCl_2$-$BaI_2$ salt can be easily removed by simply dissolving the salt in water at room temperatures due to their high solubility in water (~ 140 g KI / 100 mL water, 35.8 g $BaCl_2$ / 100 mL water, and 221 g $BaI_2$ / 100 mL water,). The KI flux removal and crystal extraction steps are typically completed within one or two minutes, while for $BaCl_2$-$BaI_2$ mixture, these procedures usually take 10 - 15 minutes.

**Room temperature chemical and structural characterization**

Figure 2a illustrates a scanning electron microscopy (SEM) image of a representative KI flux-grown $BaTiS_3$ crystal that is about one order thicker than regular CVT-grown "thin" needles. Energy dispersive X-ray spectroscopy (EDS) measurements were conducted on the crystal to assess its chemical composition. The mapping results, as depicted in the inset of Figure 2a, show a uniform distribution of Ba, Ti, and S elements throughout the field of view of the crystal. Figure 2b plots a representative EDS spectrum that quantifies a Ba : Ti : S atomic ratio of 1 : 1.02 : 2.96, suggesting a good stoichiometry. Noteworthily, no presence of either K or I element was observed within the detection limit of EDS with a standard deviation of 0.2% in atomic percentage, which indicates that the use of KI as a salt flux did not introduce a significant number of unintended elements in BaTiS$_3$ crystals during the growth.

A representative thin-film out-of-plane X-ray diffraction (XRD) scan of such crystals is shown in Figure 2c, where sharp 010-type reflections indicate the (010)-orientation of the crystal surface. We also performed polarization-resolved Raman spectroscopy on flux-grown BaTiS$_3$ crystals to reveal the anisotropic vibrational lattice properties (Figure 2d). Moreover, scanning transmission electron microscopy (STEM) studies were conducted to further confirm the quality of flux-grown BaTiS$_3$ crystals. An atomically resolved high-angle annular dark-field (HAADF) image of a BaTiS$_3$ crystal grown by KI flux, along the *c*-axis and its corresponding fast Fourier transform (FFT), as shown in Figure 2e, clearly reveal the hexagonal arrangement of the Ba (the brightest columns) and Ti atomic columns (less bright columns). Additionally, electron energy loss (EEL) spectroscopy data was collected simultaneously for mapping the distribution of Ba, Ti, S, as illustrated in Figure S2.

**Optical anisotropy**

Optical anisotropy is one of the most exciting physical properties of BaTiS$_3$ crystals[2]. A giant broadband birefringence ($\Delta n = |n_\mathrm{e} - n_\mathrm{o}|$, where $n_\mathrm{e}$ and $n_\mathrm{o}$ are the extraordinary and ordinary refractive index respectively) of up to 0.76 was reported in the mid- to long-wave infrared region in 2018[6], which became a record at the time and is still among the highest in recently reported anisotropic optical crystals[24]. Moreover, two distinctive absorption edges at 0.28 eV and 0.78 eV were observed in CVT-grown BaTiS$_3$ crystals, leading to a large dichroic window in between[2].

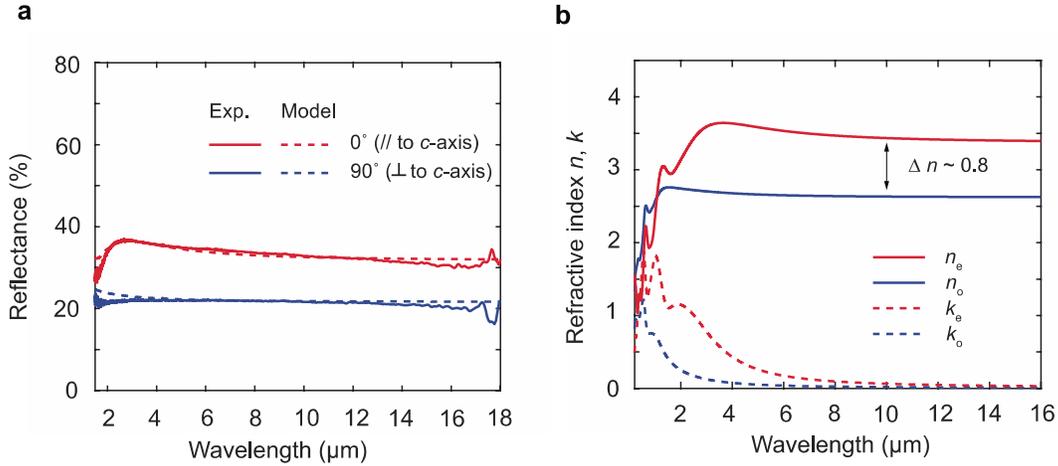

**Figure 3 Optical anisotropy of flux-grown BaTiS$_3$.** (**a**) Infrared reflectance spectra of a KI flux-grown BaTiS$_3$ crystal with incident light polarized at 0° and 90°, with respect to the *c*-axis. (**b**) Extracted complex refractive-index values of flux-grown BaTiS$_3$ for the ordinary (perpendicular to *c*-axis) and extraordinary (parallel to *c*-axis) from visible to mid-infrared region.

We first evaluated the in-plane optical anisotropy of KI flux-grown BaTiS$_3$ crystals using polarization-resolved reflectance and transmittance using Fourier-transform infrared spectroscopy (FTIR). Figure 3a shows the reflectance spectra up to 18 μm that were collected with incident light polarized at 0° and 90° with respect to the *c*-axis, on a freestanding BaTiS$_3$ crystal with *a*- and *c*-axes in-plane (~ 250 μm thick). The large difference in the reflectance values between different polarizations clearly signifies that the flux-grown BaTiS$_3$ crystals are optically anisotropic similar to past reports[2,19]. Unlike the optical spectra of thin platelets of BaTiS$_3$ crystals grown by the CVT method[2], the Fabry–Pérot fringes are absent from this measurement due to the substantial thickness of this flux-grown crystal.

To quantify the optical anisotropy, we extracted the complex refractive index for wavelengths from 210 nm to 16 μm by combining the FTIR measurements and spectroscopic ellipsometry measurements from 210 nm to 2.5 μm, following the analysis procedures reported elsewhere[2,24]. Detailed fitting and analyses are presented in Methods and Supplementary Information. Figure 3b plots the corresponding dispersion of the real ($n$) and imaginary part ($\kappa$) of the refractive index for the ordinary (perpendicular to c-axis) and extraordinary (parallel to c-axis) directions. In the region with wavelengths of ~ 4 µm and above, the flux-grown BaTiS$_3$ shows a large birefringence of ~ 0.8, which is consistent with the reported values[2]. The model fits well for the reflectance spectra, as shown in Figure 3a.

However, the extracted dispersion of $\kappa$ values may not be very accurate due to the absence of Fabry–Pérot fringes and the lack of information on absorption edges in the transmission spectra of thick samples (Figure S8). We determined the two absorption edges of flux-grown BaTiS$_3$ at ~0.35 eV and ~0.77 eV, respectively, from the transmission spectra of a thinner piece of crystal (~ 40 µm thick), which clearly reveals a large dichroic window and a slight shift of the low-energy edge, compared to reported values measured on CVT-grown crystals[2], as shown in Figure S9. The absence of substantial difference in transmittance spectra between different polarizations and the low absolute values of transmittance in thick flux-grown BaTiS$_3$ crystals (Figure S8) can be potentially attributed to the non-negligible scattering due to the defects in the crystals, misalignment of crystal optical axes or the use of an objective lens (NA of 0.17 and 0.4 in our measurements). Further improvements in the crystal quality, sizes and the employment of large beam spot sizes during optical characterization shall be able to help resolve the issue. Nonetheless, the optical properties of flux-grown BaTiS$_3$ crystals are found to be largely consistent with those measured on CVT-grown samples[2], featuring both giant birefringence values in transparent regions and a large dichroism window.

**Phase transitions**

Besides its intriguing optical properties, BaTiS$_3$ has gained substantial research attention recently due to the discovery of electronic phase transitions[8] and the associated electronic functionalities[9] such as resistive switching. Several important questions concerning the underlying mechanism of phase transitions in such a semiconducting system were also raised[8]. An in-depth understanding of the CDW phenomena in BaTiS$_3$ necessitates experimental investigations into its phonon dispersion and electronic structure, using advanced techniques such as inelastic neutron scattering and angle-resolved photoemission spectroscopy. However, due to the limitations on the crystal dimensions of CVT-grown BaTiS$_3$ crystals, many of these techniques cannot be employed. Therefore, the flux growth method presented in this study, along with further growth optimizations, offers a promising route to overcome these limitations in characterizing BaTiS$_3$, towards unraveling the phase transition mechanism.

Here, we employed temperature-dependent structural and transport measurements to probe these phase transitions. Single crystal XRD measurements were carried out on a BaTiS$_3$ crystal grown by KI flux at 100 K, 240 K and 300 K, respectively during the warming cycle, to reveal three different phases across the two phase transitions in BaTiS$_3$. Figure 4a to 4c show the precession maps of flux-grown BaTiS$_3$ crystals projected onto the *hk*2 reciprocal plane at the corresponding temperatures, and Table S1 summarizes the evolution of unit cell sizes across the phase transitions, both of which agree well with previous reports of CVT-grown BaTiS$_3$ crystals[8]. Within the CDW phase, a series of weak superlattice reflections in diffraction patterns emerged compared to the room temperature structure, which is associated with a unit cell doubling in *a-b* plane across the charge density wave transition; while at 100 K, the superlattice peaks disappeared and the structure was completely distinctive from the room temperature phase. Similar results showing all three distinctive phases were observed in (001)-oriented BaTiS$_3$ crystals grown by the BaCl$_2$-BaI$_2$ flux, and the corresponding precession maps are illustrated in Figure S4.

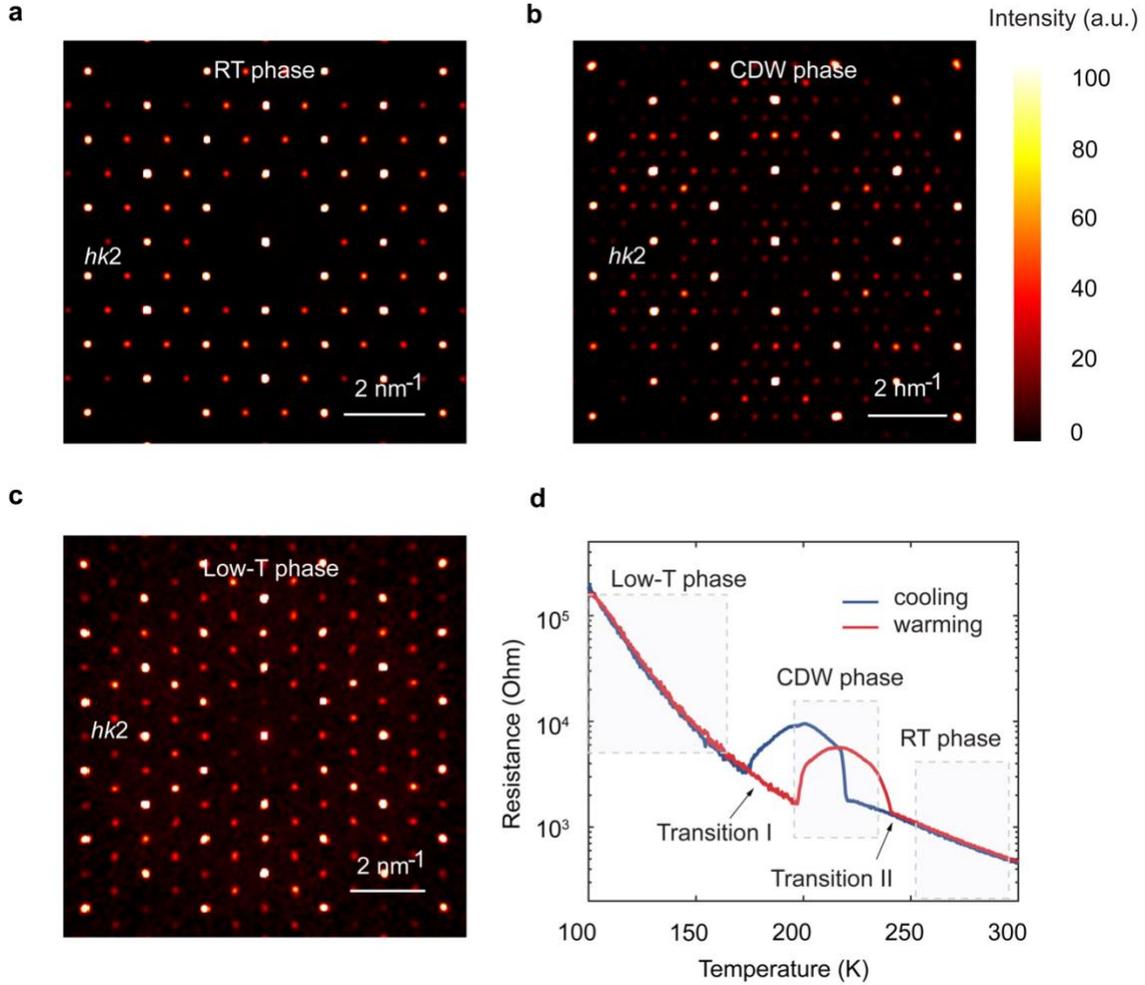

**Figure 4 Phase transitions in flux-grown BaTiS$_3$.** (**a**) to (**c**) Reciprocal precession images of flux-grown BaTiS$_3$ crystal along *hk*2 projection for three different phases. (**d**) Representative temperature dependent electrical resistance of flux-grown BaTiS$_3$ crystal along the *c*-axis. Both the hysteretic phase transitions exist, consistent with CVT-BaTiS$_3$.

Further, we conducted electrical transport measurements on a flux-grown BaTiS$_3$ to study the electronic phase transitions, as shown in Figure 4d. A relatively thin plate-like crystal (~ 50 µm thick) grown by BaCl$_2$-BaI$_2$ flux was intentionally selected for the ease of device fabrication, following the procedures reported elsewhere[8,25]. Here, we identified transport anomalies with hysteresis windows at 175 – 200 K (Transition I) and 220 – 240 K (Transition II), respectively, which correspond to the two phase transitions reported in BaTiS$_3$. A slight shift of the Transition II (~ 20 K) towards the lower temperature and a reduced thermal hysteresis window of Transition I were observed in flux-grown BaTiS$_3$ crystals, compared to CVT-BaTiS$_3$[8], despite the overall consistency in transport behavior. Although the origin of this discrepancy is not fully understood, we attribute it to the differences in the specific crystal growth conditions, considering the susceptibility of such phase transitions to a range of extrinsic factors such as strain[8] and potentially defects[26]. Nevertheless, this study is expected to promote in-depth investigations into the growth of large single crystals of complex chalcogenides, such as BaTiS$_3$, using flux growth and other techniques such as solution Bridgman growth method.

## Conclusions

In conclusion, we have successfully synthesized large-scale single crystals of BaTiS$_3$ utilizing a molten flux method with either KI or BaCl$_2$-BaI$_2$ as the salt flux, from which the crystals grow an order of magnitude larger in sizes than those prepared using the conventional CVT approach. Importantly, this growth method has preserved both the giant optical anisotropy and electronic phase transitions observed in crystals grown by vapor transport method. This confirms the high quality of these flux-grown BaTiS$_3$ crystals. Our study of flux growth of BaTiS$_3$ opens up new opportunities for its practical device applications in optics and optoelectronics, and it facilitates advanced material characterization towards a better understanding of these intriguing physical phenomena.

Moreover, we anticipate this flux-based crystal growth method, particularly with KI as the salt flux, would be broadly applicable for the synthesis of other emerging complex chalcogenide materials, such as highly anisotropic Sr$_{1+x}$TiS$_3$ that was recently reported to show a new record high birefringence of 2.1[24] for optical applications and BaZrS$_3$ for potential photovoltaic applications[1,27]. This method offers advantages over the utilization of highly corrosive BaCl$_2$ from various perspectives, as discussed in this manuscript. For BaCl$_2$-BaI$_2$ flux, the reduced melting temperature of the flux mixture upon the addition of BaI$_2$ is also expected to favor the growth of large crystals. Nonetheless, it is important to note that unintended potassium doping, or iodine incorporation may pose challenges in certain specific material systems, which would need careful chemical analysis to clarify. Further, the successful demonstration of molten flux growth can enable more advanced growth techniques such as solution Bridgman growth, towards the synthesis of crystals with sizes that are suitable for neutron studies, or even as single crystal chalcogenide wafers, which are important for in-depth material characterization and achieving high-quality epitaxial thin film growth of complex chalcogenides in the future.

## Methods

**Crystal synthesis**

Polycrystalline BaTiS$_3$ powder was synthesized by first mixing a stoichiometric amount of barium sulfide powder (Sigma-Aldrich, 99.9%), titanium powder (Alfa Aesar, 99.9%), with a 5% excess of sulfur (Sigma-Aldrich, 99.998%). The mixture was loaded into a sealed, evacuated quartz ampoule and heated to 1040°C at a rate of 50°C/h. The ampoule was cooked for 160 h and then quenched in NaCl / ice bath. The resulting material is in the form of loosely packed black powder. The obtained BaTiS$_3$ powder was then loaded into another quartz ampoule with sulfur pieces (BaTiS$_3$ : S weight ratio ~ 2:1) for sulfur annealing. The ampoule was cooked at 650°C for 80 hr in a tube furnace. The excess sulfur was separated from BaTiS$_3$ powder by re-heating the mixture such at sulfur could condense at the cold end.

For single-crystal growth using KI, 10 g pre-dried KI powder (Alfa Aesar, 99.9%) and 200 mg pre-synthesized BaTiS$_3$ powder (BaTiS$_3$:KI atomic ratio ~ 1:100) were placed in a quartz ampoule with 19 mm of outer diameter (OD), 2 mm of wall thickness, and approximately 8 cm in length, mixed with and 10 mg excess S powder in a N$_2$-filled glovebox. The ampoule was then sealed under vacuum. It is important to dry the as-purchased KI powder in glovebox at 110°C overnight before using, in order to avoid explosion at high temperatures. For BaTiS$_3$ crystal growth using BaCl$_2$-BaI$_2$ mixture, 0.5 BaCl$_2$ powder (Alfa Aesar, 99.998%), 1.5 g BaI$_2$ powder (Alfa Aesar, 99.999%), and 50 mg pre-synthesized BaTiS$_3$ powder were thoroughly mixed and placed in a small aluminum crucible (Canfield crucible set, 2 mL), before loading into a straight quartz ampoule (15 mm inner diameter (ID), 19 mm OD, ~20 cm long). Quartz wool was used to fix the aluminum crucible position and a small quartz cap (10 mm ID, 14 mm OD, ~ 3 cm long) was used to seal the ampoule. Figure S5 shows an optical image of a sealed ampoule for BaCl2-BaI2 flux growth.

After sealing, the quartz ampoules were placed in an alumina crucible and then loaded into a box furnace (modified from a 6-inch three zone tube furnace) with a vertical configuration as illustrated in Figure 1a. The ampoules were heated to 1040°C in 30 h and hold at 1040°C for 30 h, followed by a slow cooling step to 700°C at 1°C/h before a natural cooling down process (Figure 1c). After about 20 days, the ampoule was removed from the furnace and washed with DI water to remove the excess salt. Crystals were picked individually after drying under an optical stereo microscope for further characterizations. Figure S6 illustrates optical images of several different BaTiS$_3$ crystal synthesis steps using KI flux.

**Chemical, vibrational and electron microscopic characterization**

Energy dispersive X-ray spectroscopy (EDS) was performed using UltimMax 170 spectrometer attached on a NanoSEM 450 system. Polarization-dependent Raman spectroscopy was performed in a backscattering geometry using a conformal microscopic spectrometer. The incident beam (532 nm) was linearly polarized, and a half-wave plate was used for polarization rotation.

Scanning transmission electron microscopy (STEM) imaging was performed using an aberration corrected Nion UltraSTEM 100 (operated at 100 kV) microscope at Oak Ridge National Laboratory. HAADF-STEM images were acquired using a convergence semi-angle of 30 mrad and an annular dark-field detector with inner and outer collection semi-angles of 80 and 200 mrad, respectively. EELS was carried out using a Gatan Enfina EEL spectrometer attached to the Nion UltraSTEM. A collection semi-angle of 48 mrad and an energy dispersion of 1 eV per channel was used to acquire EELS data.

**Infrared spectroscopy and ellipsometry**

Polarization-resolved infrared spectroscopy was conducted using a Fourier transform infrared spectrometer (Bruker Optics Vertex 70) and an infrared microscope (Hyperion 2000). A 15× Cassegrain microscope objective (NA = 0.4) was used for both transmittance and reflectance measurements at near-normal incidence on a (100) face of a KI-grown $BaTiS_3$ crystal. The polarization of incident light was controlled using a wire-grid polarizer. The measurements were performed using a Globar source, a potassium bromide beam splitter and a mercury cadmium telluride detector.

Ellipsometry measurements were carried out using a VASE ellipsometer (J. A. Woollam Co.) with focusing probes over a spectral range of 210 nm to 2.5 µm at an angle of incidence of 55˚. Data were acquired from three different sample orientations (optical axis parallel, perpendicular, and ~48˚ to the plane of incidence). Data analysis and refractive index extraction were performed using WVASE software (J. A. Woollam Co.). See more details in the Supplementary Note 1.

**X-ray diffraction**

The single crystal X-ray diffraction measurements were carried out on a Rigaku XtaLAB Synergy-S diffractometer at University of Southern California. Crystals were mounted on MiTeGen Dual Thickness MicroMounts[TM] and placed in a nitrogen cold stream on the goniometer-head of the diffractometer, which is equipped with a Mo K$\alpha$ X-ray source (wavelength 0.71Å) and a HyPix-6000HE detector. Diffraction data were collected ensuring at least 99.9% completeness for a resolution of 0.70 Å. Data reduction, scaling, unit cell determination, and precession map analysis were done in CrysAlisPro.

The thin-film out-of-plane XRD scan was performed in a Bruker D8 Advance diffractometer using a Ge (004) two bounce monochromator with Cu K$_{\alpha 1}$ radiation at room temperature.

**Device fabrication and transport measurements**

Flux-grown BaTiS$_3$ crystal was first embedded in a polyimide medium for planarization[25] and then regular photolithography and ebeam evaporation were applied to form electrodes, following the procedures reported elsewhere. Standard low-frequency ($f$ = 17 Hz) AC lock-in techniques were used to measure sample resistance in four-probe geometry from 100 K to 300 K, with an excitation current of about 100 nA.

# Author declarations

**Author contributions**

**Acknowledgements**

This work was supported by an ARO MURI with award number W911NF-21-1-0327, an ARO grant with award number W911NF-19-1-0137, an NSF grant with award numbers DMR-2122070 and DMR-2122071, and an ONR grant with award number N00014-23-1-2818. Scanning transmission electron microscopy was supported by the Center for Nanophase Materials Science (CNMS), which is a US Department of Energy, Office of Science User Facility at Oak Ridge National Laboratory. An NSF grant with award number CHE-2018740 provided funds to acquire the Rigaku XtaLAB Synergy-S diffractometer that was used for single-crystal XRD studies. The authors gratefully acknowledge the use of facilities at John O'Brien Nanofabrication Laboratory and Core Center for Excellence in Nano Imaging at University of Southern California for the results reported in this manuscript. H.M. and M.K. acknowledge the support from the Office of Naval Research (N00014-20-1-2297). The authors also acknowledge the use of facilities and instrumentation at the UW-Madison Wisconsin Centers for Nanoscale Technology (wcnt.wisc.edu) partially supported by the NSF through the University of Wisconsin Materials Research Science and Engineering Center (DMR-1720415).

**Conflict of interest**

The authors declare no competing financial interests.

# Supplemental Information for

# Molten flux growth of single crystals of quasi-1D hexagonal chalcogenide BaTiS$_3$

**Authors:** Huandong Chen[1], Shantanu Singh[1], Hongyan Mei[2], Guodong Ren[3], Boyang Zhao[1], Mythilli Surendran[1,4], Yan-Ting Wang[1], Rohan Mishra[5,3], Mikhail A. Kats[2,6], Jayakanth Ravichandran[1,4,7*]

[1]Mork Family Department of Chemical Engineering and Materials Science, University of Southern California, Los Angeles CA, USA

[2]Department of Electrical and Computer Engineering, University of Wisconsin-Madison, Madison WI, USA

[3]Institute of Materials Science & Engineering, Washington University in St. Louis, St. Louis MO, USA

[4]Core Center for Excellence in Nano Imaging, University of Southern California, Los Angeles CA, USA

[5]Department of Mechanical Engineering and Materials Science, Washington University in St. Louis, St. Louis MO, USA

[6]Department of Materials Science and Engineering, University of Wisconsin-Madison, Madison WI, USA

[7]Ming Hsieh Department of Electrical and Computer Engineering, University of Southern California, Los Angeles CA, USA

*Email: j.ravichandran@usc.edu

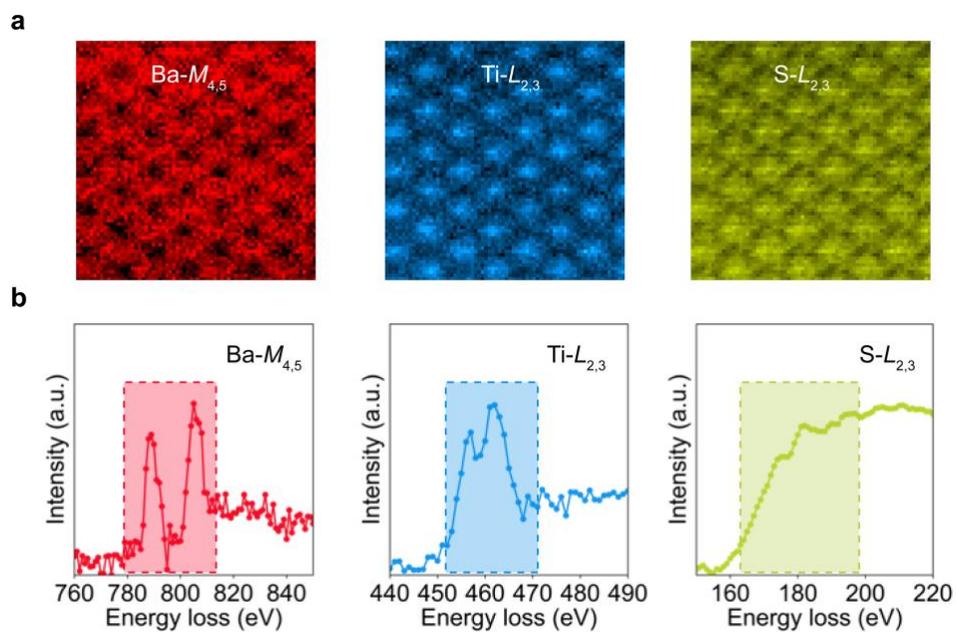

**Figure S1.** (**a**) Elemental edge maps using Ba-*M*, Ti-*L*, and S-*L* edges extracted from an EELS image. (**b**) The EEL spectra of the corresponding elements used to make the elemental maps in (**a**). The energy range used for signal integration have been highlighted with shaded boxes.

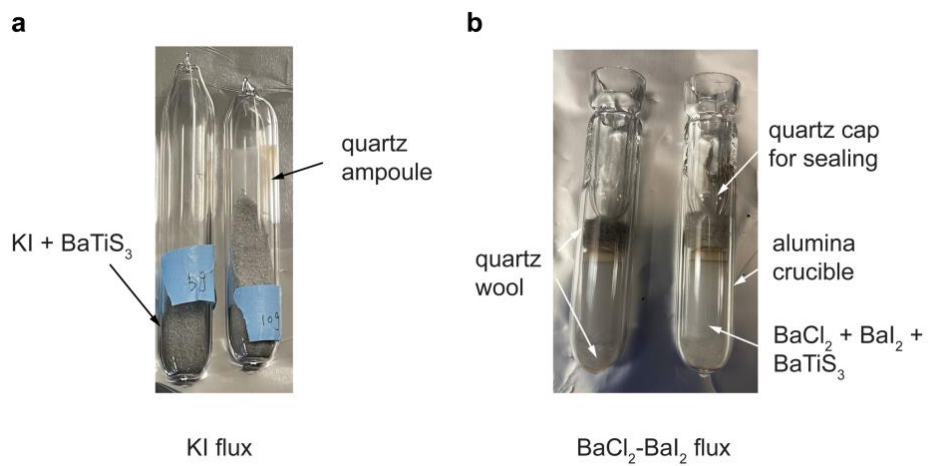

**Figure S2.** Optical images of quartz ampoules prepared for (**a**) KI flux growth, and (**b**) $BaCl_2$-$BaI_2$ flux growth. Additional alumina crucible was used as inner container for $BaCl_2$-$BaI_2$ flux to minimize its corrosion to quartz ampule.

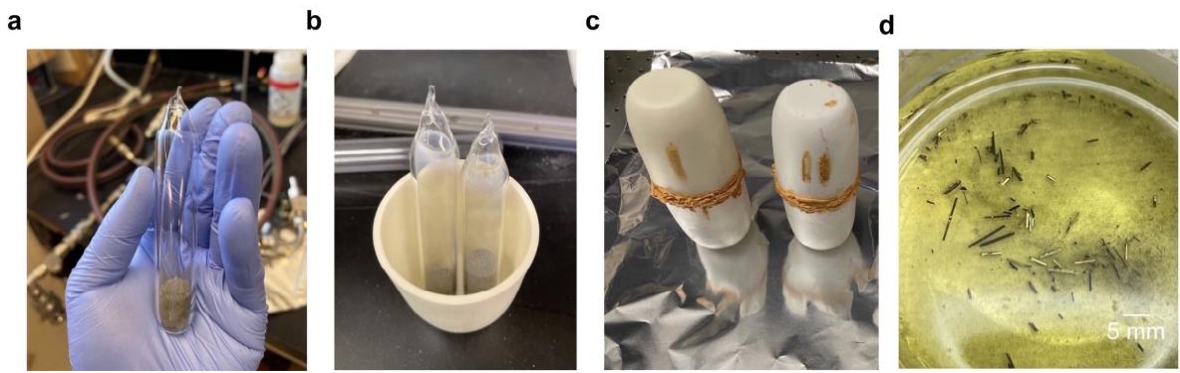

**Figure S3.** Optical images of several different steps of BaTiS$_3$ crystal synthesis using KI flux: (**a**) after sealing KI flux, pre-synthesized BaTiS$_3$ powder and excess sulfur in a quartz ampule, (**b**) after loading samples in an alumina crucible, (**c**) after sealing alumina crucibles using high temperature cement, this step protects the furnace from any potential leaking, and (**d**) removing excess KI flux in a crystallizing dish using DI water.

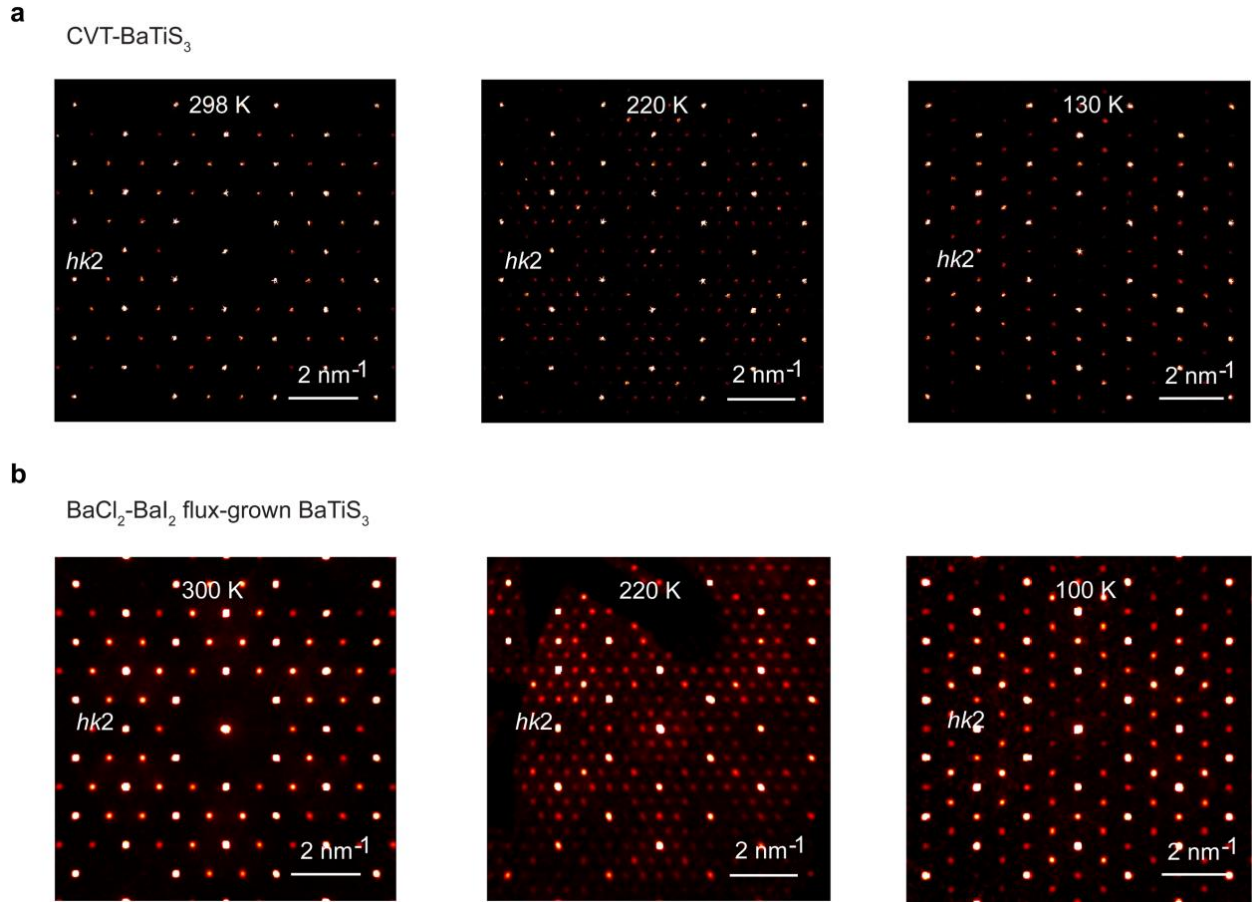

**Figure S4.** (**a**) Precession maps of a conventional CVT-grown BaTiS$_3$ crystal along $hk2$ projection for three different phases. The data was adapted from the Reference[1]. (**b**) Reciprocal precession images of a BaCl$_2$-BaI$_2$ flux-grown BaTiS$_3$ crystal along the same projection, taken at 300 K, 220 K, and 100 K, respectively during the warming cycle.

**Table S1. Evolution of flux-grown BaTiS$_3$ unit cell sizes at different temperatures**

| Flux-BaTiS$_3$ | 300 K | 240 K | 100 K |
|---|---|---|---|
| $a$, $b$, $c$ (Å) | 11.71, 11.71, 5.84 | 23.38, 23.38, 5.85 | 13.46, 13.46, 5.83 |
| α, β, γ (°) | 90, 90, 120 | 90, 90, 120 | 90, 90, 120 |
| space group | $P6_3cm$ | $P3c1$ | $P2_1$ |

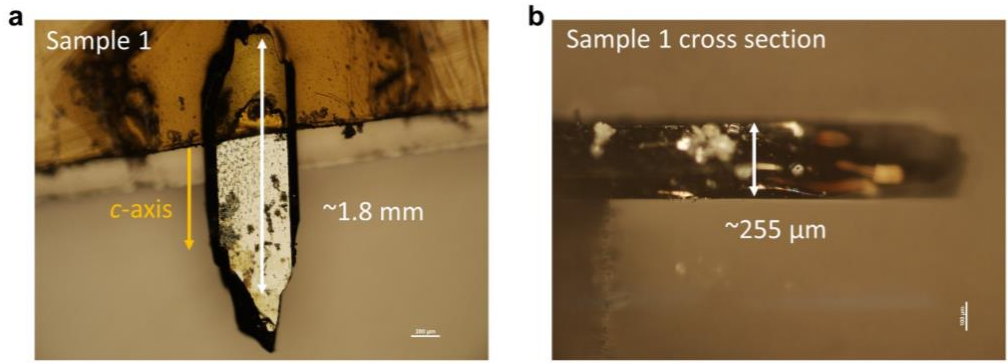

**Figure S5.** (**a**) optical images of a KI flux grown BaTiS$_3$ crystal plate, with dimension along the *c*-axis larger than 1.8 mm. (**b**) cross-section image of a KI flux grown BaTiS$_3$ crystal plate, showing the thickness of ~255 μm.

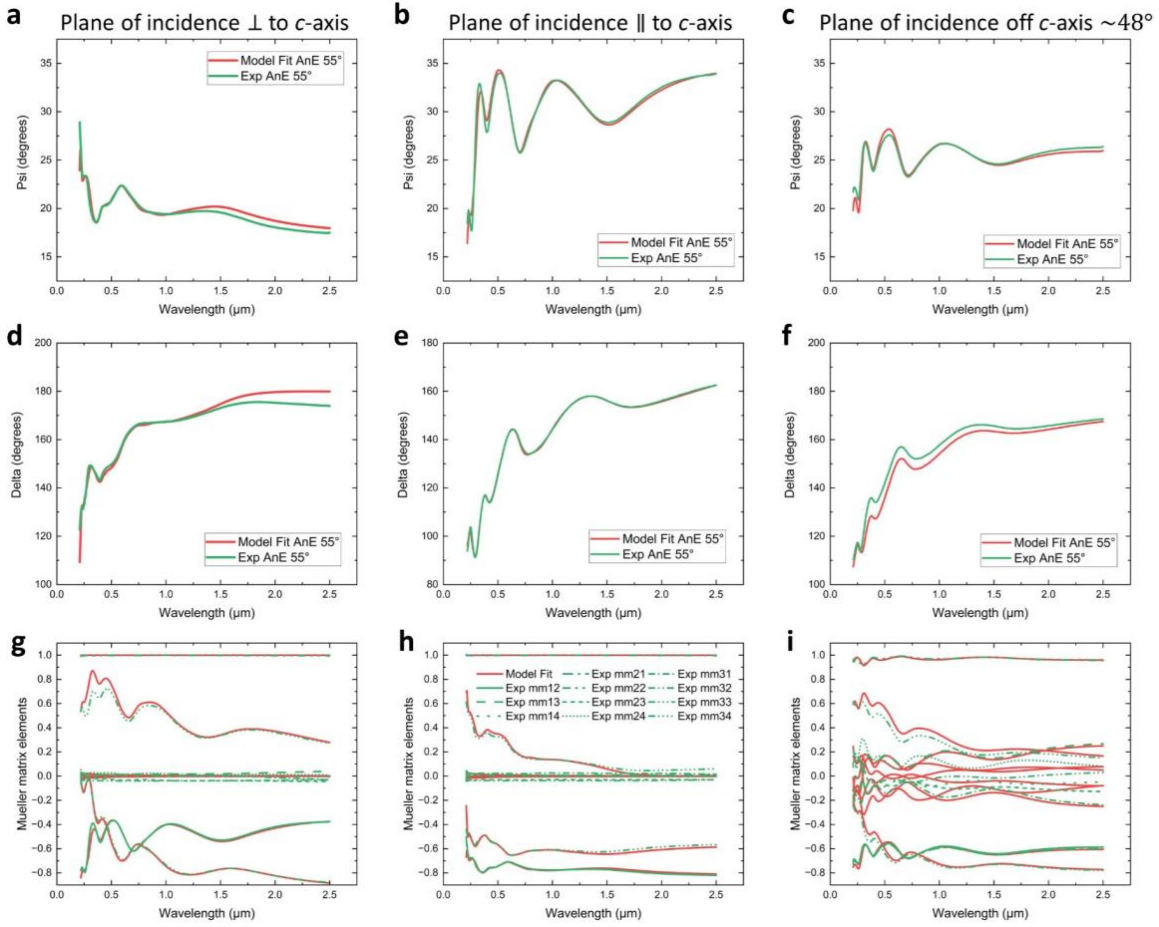

**Figure S6.** Raw ellipsometry data (Ψ **(a-c)** and Δ **(d-f)** of $A_{ne}$, Mueller matrix elements **(g-i)**) for KI flux grown BaTiS$_3$ at $\theta_i = 55°$ with the crystal *c*-axis perpendicular, parallel, and at an off-axis angle ~48° to the plane of incidence, showing good consistence between the experimental data (green lines) and model fit data (red lines).

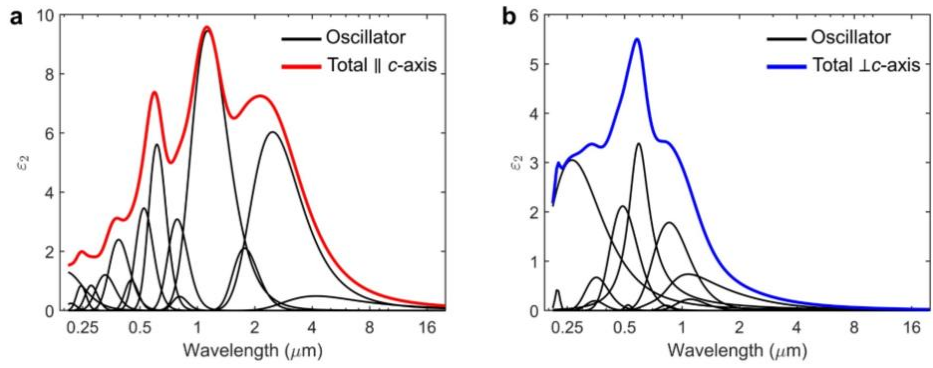

**Figure S7.** The $\varepsilon_2$ values calculated from the final optical oscillator model used to fit to the ellipsometry data of KI flux grown $BaTiS_3$. The optical oscillator model is built on Kramers-Kronig consistent oscillators.

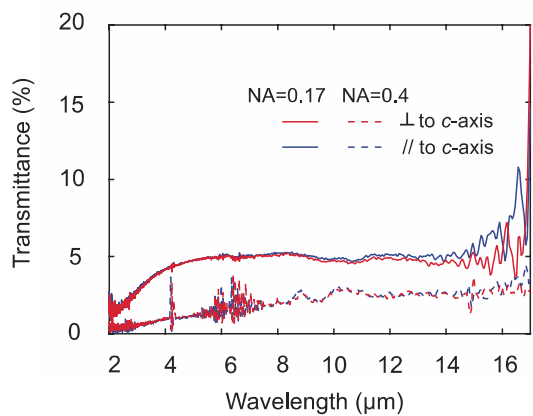

**Figure S8.** Infrared transmission spectra of a thick KI flux-grown BaTiS$_3$ crystal with incident light polarized at 0° and 90°, with respective to the *c*-axis. No substantial difference was observed between different polarizations and the absolute values of transmittances are low. Two different objective lenses (15×, NA = 0.4, Casegrain reflective objective; 5×, NA = 0.17, Ge IR refractive objective) have been used for the measurements. The corresponding reflectance spectra with NA = 0.4 is shown in Figure 3a.

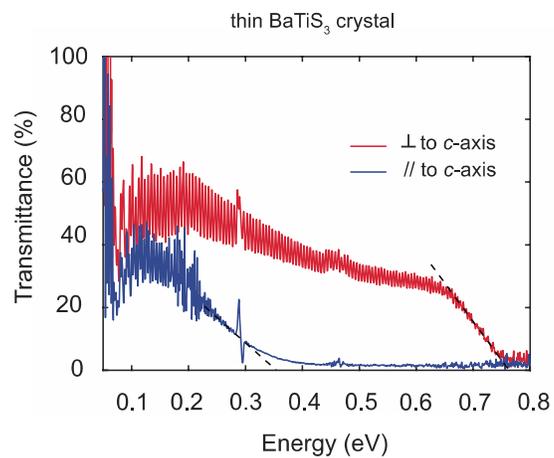

**Figure S9.** Infrared transmission spectra of a relatively thin KI flux-grown BaTiS$_3$ crystal (~ 40 µm thick) with incident light polarized perpendicular and parallel to the *c*-axis. Two absorption edges at ~ 0.35 eV and ~ 0.77 eV are clearly revealed. Energy scale was used for the ease of analysis.

**Supplementary Note 1. Extraction of the Anisotropic Complex Refractive Index**

Generalized spectroscopic ellipsometry (VASE + focusing optics, J.A. Woollam Co.) and polarized reflection/transmission infrared spectroscopy (Bruker VERTEX 70 + Hyperion 2000 with MCT IR detector) were employed to extract the complex refractive index tensor of KI flux grown $BaTiS_3$ from 210 nm to 17 µm. The $BaTiS_3$ crystal was suspended over air with no substrate during the measurement, as shown in Figure S5.

Four ellipsometry measurements were made under Mueller Matrix Mode with three different crystal orientations: with the crystal $c$-axis perpendicular to the plane of incidence; parallel to the plane of incidence; and at off-axis angles (~48°) and the three measurements were made with an angle of incidence of 55° from the surface normal.

Figure S6 shows the fitting between the raw measured ellipsometry data (in green) and the corresponding model data (in red) for the three measurements of KI flux grown $BaTiS_3$, consisting of the generalized ellipsometry data types $A_{ne}$ and Mueller matrix elements $mm_{12}, \ldots, mm_{34}$. $A_{ne}$ is the ratio between the two Fresnel coefficients $r_{pp}$ and $r_{ss}$. Mueller matrix elements describe how the sample transforms the Stokes vector of the incident light. The $m_{11}$ element represents the total intensity of the reflection or transmission of the sample and the VASE ellipsometer records other Mueller matrix elements normalized to $m_{11}$, so the quantities range between -1 and +1[2].

$$A_{ne} = \rho = \frac{r_{pp}}{r_{ss}} = \tan(\Psi_{A_{ne}}) e^{i\Delta_{A_{ne}}}$$

$$S = \begin{bmatrix} I \\ Q \\ U \\ V \end{bmatrix} = M \begin{bmatrix} I^{in} \\ Q^{in} \\ U^{in} \\ V^{in} \end{bmatrix}, \quad M = \begin{bmatrix} m_{11} & m_{12} & m_{13} & m_{14} \\ m_{21} & m_{22} & m_{23} & m_{24} \\ m_{31} & m_{32} & m_{33} & m_{34} \\ m_{41} & m_{42} & m_{43} & m_{44} \end{bmatrix} = m_{11} \begin{bmatrix} 1 & m_{12}' & m_{13}' & m_{14}' \\ m_{21}' & m_{22}' & m_{23}' & m_{24}' \\ m_{31}' & m_{32}' & m_{33}' & m_{34}' \\ m_{41}' & m_{42}' & m_{43}' & m_{44}' \end{bmatrix}$$

All three datasets were fitted simultaneously using a single optical model with uniaxial birefringence. Here, optical properties perpendicular and parallel to the $c$-axis contain independent oscillators. By integrating the polarization-resolved reflectance of KI flux grown $BaTiS_3$ into the ellipsometry data, the optical model could be extended up to the detection limit of our FTIR (17 μm). Thus, three ellipsometry measurements and two reflectance measurements could all be fit simultaneously to a single anisotropic optical model to yield a full set of refractive indices spanning 210 nm through 17 μm. As shown in Figure S6, the measured data match well with the model fitted data.

The total oscillator model for each direction is the sum of individual Kramers-Kronig-consistent oscillators as shown in Figure S7. Table S2 lists the oscillators used in the anisotropic optical model for both the ordinary and extraordinary directions.

**Table S2. Oscillators and the fitted parameters of KI flux grown BaTiS$_3$.**

| Oscillator | Parameters (eV) | | |
|---|---|---|---|
| Gaussian<br>$\varepsilon_{Gaus} = \varepsilon_1 + i\varepsilon_2$<br><br>$\varepsilon_2 = A_n e^{-\left(\frac{E-E_n}{\sigma}\right)^2} - A_n e^{-\left(\frac{E+E_n}{\sigma}\right)^2}$<br><br>$\varepsilon_1 = \frac{2}{\pi} P \int_0^\infty \frac{\xi \varepsilon_2(\xi)}{\xi^2 - E^2} d\xi$<br><br>Where $\sigma = \frac{Br_n}{2\sqrt{\ln(2)}}$, the $1/2\sqrt{\ln(2)}$ factor defines $Br_n$ = FWHM. $P$ is the Cauchy Principal Value. | Ordinary<br>$A_1$ = 0.22881<br>$A_2$ = 1.7844<br>$A_3$ = 0.10508<br>$A_4$ = 0.11355<br>$A_5$ = 2.117<br>$A_6$ = 0.67282<br>$A_7$ = 0.19641<br>$A_8$ = 3.0509<br>$A_9$ = 0.44297<br><br>Extraordinary<br>$A_1$ = 0.49005<br>$A_2$ = 6.0344<br>$A_3$ = 2.1032<br>$A_4$ = 9.455<br>$A_5$ = 0.47327<br>$A_6$ = 3.0859<br>$A_7$ = 5.616<br>$A_8$ = 3.4602<br>$A_9$ = 1.0241<br>$A_{10}$ = 2.3982<br>$A_{11}$ = 1.2056<br>$A_{12}$ = 0.85088<br>$A_{13}$ = 0.84736<br>$A_{14}$ = 0.24468<br>$A_{15}$ = 1.3047 | Ordinary<br>$E_1$ = 1.1236<br>$E_2$ = 1.4476<br>$E_3$ = 1.5034<br>$E_4$ = 2.3794<br>$E_5$ = 2.5385<br>$E_6$ = 3.4884<br>$E_7$ = 3.5532<br>$E_8$ = 4.6747<br>$E_9$ = 5.5807<br><br>Extraordinary<br>$E_1$ = 0.29574<br>$E_2$ = 0.5<br>$E_3$ = 0.70144<br>$E_4$ = 1.0952<br>$E_5$ = 1.5367<br>$E_6$ = 1.5826<br>$E_7$ = 2.2017<br>$E_8$ = 2.3637<br>$E_9$ = 2.7584<br>$E_{10}$ = 3.2084<br>$E_{11}$ = 3.7787<br>$E_{12}$ = 4.4721<br>$E_{13}$ = 5.0477<br>$E_{14}$ = 5.7131<br>$E_{15}$ = 6.0456 | Ordinary<br>$Br_1$ = 0.4612<br>$Br_2$ = 0.73465<br>$Br_3$ = 0.27766<br>$Br_4$ = 0.23803<br>$Br_5$ = 0.97834<br>$Br_6$ = 1.1235<br>$Br_7$ = 0.91346<br>$Br_8$ = 3.3703<br>$Br_9$ = 0.36085<br><br>Extraordinary<br>$Br_1$ = 0.31132<br>$Br_2$ = 0.3465<br>$Br_3$ = 0.24885<br>$Br_4$ = 0.56054<br>$Br_5$ = 0.32465<br>$Br_6$ = 0.445<br>$Br_7$ = 0.49401<br>$Br_8$ = 0.5914<br>$Br_9$ = 0.5183<br>$Br_{10}$ = 0.91344<br>$Br_{11}$ = 1.0342<br>$Br_{12}$ = 0.79528<br>$Br_{13}$ = 0.69215<br>$Br_{14}$ = 0.86262<br>$Br_{15}$ = 2.6429 |
| Lorentz<br>$\varepsilon_L = \varepsilon_1 + i\varepsilon_2$<br><br>$\varepsilon_2 = \frac{A_n Br_n E_n}{E_n^2 - E^2 - iBr_n E}$<br><br>$\varepsilon_1 = \frac{2}{\pi} P \int_0^\infty \frac{\xi \varepsilon_2(\xi)}{\xi^2 - E^2} d\xi$<br>Where P is the Cauchy Principal Value. | Ordinary<br>$A_1$ = 3.3693 | $E_1$ = 2.1165 | $Br_1$ = 0.64235 |
| Parametric Semiconductor<br>Psemi-(M0, M1, M2, M3 & Tri)<br><br>See WVASE manual by J.A. Woollam Co.[2] | Ordinary (Psemi-Tri)<br>$A_1$ = 1.7299<br>$WL_1$ = 1.0042<br>$AL_1$ = 0.86247 | $Ec_1$ = 1.233<br>$WR_1$ = 1.0207<br>$AR_1$ = 0.1880 | $B_1$ = 0.18861 |